\documentstyle[aps]{revtex}

\newlength{\defaultparindent}
\begin{document}

\begin{center}
${\bf Electron\;impact\;excitation\;cross\;sections\;for\;allowed%
\;transitions\;in\;atoms}$

{\bf \ }

V. Fisher, \ V. Bernshtam, \ H. Golten$^1,$ \ and\ Y. Maron

{\it Faculty of Physics, Weizmann Institute of Science, Rehovot 76100, Israel%
}

$^1${\it Churchill College, Cambridge, UK }
\end{center}

\begin{quotation}
We present a semiempirical Gaunt factor for widely used Van Regemorter
formula [Astrophys. J. {\bf 136}, 906 (1962)] for the case of allowed
transitions in atoms with the $LS$ coupling scheme. Cross sections
calculated using this Gaunt factor agree with measured cross sections to
within the experimental error.
\end{quotation}

\section{Introduction}

\ Interpretation of spectroscopic measurements and simulation of kinetic and
transport processes in non-LTE plasmas require knowledge of many electron
impact excitation cross sections for atoms and ions. In general, any
excitation cross section may be calculated using one of computer codes
designed for this purpose (see, for example, Refs. [1-7]). Hundreds of cross
sections are already calculated or determined experimentally for some
intervals of incident electron energy. These results can be found in atomic
data bases.\ However, published cross sections are not usually sufficient
for detailed simulation of experiments, since data on many cross sections
are missing or do not cover the entire energy range required for calculation
of excitation rates, especially for non-Maxwellian plasmas.

In such situation, it is desirable to have easy-to-use formulae of known
accuracy applicable to various classes of transitions. Estimates of electron
impact excitation cross sections are frequently based on the Van Regemorter
formula [8-10]

\begin{equation}
\sigma _{qq^{\prime }\;}^{exc}(x)=\frac{8\pi }{\sqrt{3}}\pi
a_o^2\;f_{qq^{\prime }}\;\frac{Ry^2}{E_{qq^{\prime }}^2}\;\frac{%
G_{qq^{\prime }}(x)}x\;\;,  \label{1}
\end{equation}
which is derived for single-electron electric dipole transitions (in other
words, for optically allowed transitions) [11]. Here $\sigma _{qq^{\prime
}\;}^{exc}(x)\;$is the electron-induced excitation cross section from the
lower state $q\;$into the upper state $q^{\prime }$, $\ x=\varepsilon
/E_{qq^{\prime }}$, $\varepsilon \;$is the kinetic energy of relative motion
between projectile electron and target atom (ion), $E_{qq^{\prime }}$ is the
transition energy, $a_o$ is the Bohr radius, $Ry$ is the Rydberg energy
unit, $f_{qq^{\prime }}\;$is the absorption oscillator strength, and $%
G_{qq^{\prime }}(x)\;$is the Gaunt factor (which may be treated as a fitting
function of order unity). It is known that formula (1) provides a better fit
to experimental data if different expressions for the Gaunt factor are used
when applied to atoms, singly-charged ions, or multiply-charged ions; and to
transitions with $\Delta n=0$ or $\Delta n>0\;$[4,8-10,14-16]$.$ Probably,
the fit may be improved further if a dependence on other transition
parameters is introduced in the Gaunt factor, for example, dependence on the
orbital quantum number of the optical electron $\ell ,\;$as is found for
multiple ions [17] using high-accuracy theoretical results.

After the first publication by H. Van Regemorter in 1962 [8], there were a
few attempts to infer reasonably accurate Gaunt factors for various classes
of transitions including non-dipole and intercombination ones [9,10,14-16].
The Gaunt factors obtained do not provide an accuracy of about 10-30\% which
is expected from atomic codes, and there is some criticism of the use of the
Van Regemorter approximation in the epoch of computers [5,18]. Nevertheless,
the simplicity of the Van Regemorter formula makes it attractive for
estimates, and it is reasonable to improve the accuracy of this formula by
updating the Gaunt factor for various classes of transitions.

In section 2 we present rather accurate Gaunt factor for a broad and
important class, namely, for allowed transitions between $n\ell $-states in
atoms with $LS$ coupling. These transitions may be represented by the scheme 
\begin{equation}
\gamma n\ell ^m\;^{2S+1}L\;\rightarrow \gamma n\ell ^{m-1}n^{\prime }\ell
^{\prime }\;^{2S^{\prime }+1}L^{\prime }  \label{2}
\end{equation}
with selection rules [13] 
\begin{equation}
n^{^{\prime }}\geq n,\;\;\;\;\ell ^{\prime }=\ell \pm 1,\;\;\;\;\;S^{\prime
}=S,\;\;\;\;\;L^{\prime }-L=0,\pm 1\;\;\;\text{and}\;\;\;L+L^{\prime }>0;
\label{3}
\end{equation}
$\gamma \;$denotes all subshells which do not change their state in the
collision. Our results relate to the case when the excitation occurs in the
outer shell. Applicability of the Gaunt factor obtained to excitation of
inner-shell electrons is not checked because of no experimental data on the
cross sections.

\section{The Gaunt factor}

Tables 1 and 2 present a list of experimentally studied electron-induced
transitions which belong to the class considered here, namely, allowed
transitions (2),(3) in atoms. For convenience of further analysis,
transitions with $\;\Delta n=0$\ are listed separately from transitions with 
$\Delta n>0.\;$Various publications present from one to a hundred
experimental points for each of the studied cross sections. To exclude any
dominating influence of Ref. [19], we use not more than 40 values for each
cross section from any publication. In the case, when only part of the
experimental points are taken, the points taken are either every second
point or every third point along $x$ for the cross section of interest. The
number of accounted points is given in brackets in the fifth column of the
tables.

Values of the Gaunt factor inferred from the experimentally studied cross
sections $\sigma _{qq^{\prime }\;}^{exc}(x)\;$for transitions with $\Delta
n=0$ are demonstrated in Figure 1. This data may be fitted rather well by
the expression 
\begin{equation}
G_0(x)=(0.33-0.3x^{-1}+0.08x^{-2})\ln x  \label{4}
\end{equation}
shown by the solid curve. The subscript 0 denotes the condition $\Delta
n=0.\;$To illustrate the accuracy of the above expression, Figure 2 presents
a histogram of discrepancies D$_k$ between experimental values $G_k^{\exp }$
and semiempirical Gaunt factor G$_0$($x_k$). The discrepancy is defined as a
ratio 
\begin{equation}
D_k=\frac{\mid G_k^{\exp }-\;G_0(x_k)\mid }{G_0(x_k)},  \label{5}
\end{equation}
where $k$ is the order number of the experimental point, and $x_k$ is the
value of $x$ for this point. The histogram demonstrates the numbers of
experimental points per 10\% intervals of increasing D. One can see that for
95\% of experimental points the accuracy of the Gaunt factor (4) is better
than $\pm 50\%..\;$This accuracy seems to be acceptable for any estimates.
For 82\% of the points the accuracy is better than $\pm $30\%..

Values of the Gaunt factor inferred from experimentally studied cross
sections $\sigma _{qq^{\prime }\;}^{exc}(x)\;$for transitions with $\Delta
n>0$ are demonstrated in Figure 3. This data may be fitted rather well by
the expression 
\begin{equation}
G_{>}(x)=(0.276-0.18x^{-1})\ln x  \label{6}
\end{equation}
shown by the solid curve. The subscript 
\mbox{$>$}
denotes the condition $\Delta n>0.\;$When $x\rightarrow \infty ,$

\begin{equation}
G_{>}(x)\approx \frac{\sqrt{3}}{2\pi }\ln x  \label{7}
\end{equation}
and expression (1) becomes the Bethe formula [9]. For $x$ 
\mbox{$<$}
$\;10,$\ expression (6) provides a better fit to experimental data than the
asymptotic expression (7) shown by the dotted line. The accuracy of the
Gaunt factor (6) is demonstrated by the histogram in Figure 4. For 82\% of
experimental points the Gaunt factor is accurate to better than $\pm $30\%.

From 2 to$\;$3 \% of experimental points deviate from the Gaunt factor given
by expressions (4) and (6) by more than a factor of 2.\ Most of these points
belong to the energy range $x\approx 1\;$where the cross sections are small
and measurements are less accurate. It is worth mentioning that the
discrepancy between experimental results obtained by various research groups
(for the same studied transition), is of a similar magnitude to the
deviation of the experimental data from the Gaunt factor given by
expressions (4) and (6). To illustrate this fact, figures $5,6,$and $7\;$%
present experimental data obtained for three transitions: (i) {\sl 3s }${\sl %
\rightarrow }${\sl \ 3p }in{\sl \ Na, }(ii){\sl \ 4s }$\rightarrow ${\sl \
4p }in{\sl \ K, }(iii){\sl \ 1s}$^{2\;1}{\sl S\rightarrow 1s4p\;}^1{\sl P\;}$%
in${\sl \ He.\ \ \ \ \ \ \ }$

Some experimental results are not included in our analysis: results of Ref.
[53] are not included because they were shown to be inaccurate (see Ref.
[48]); results of Ref. [54] are not included because of unreliable
normalization using Ref.[53] and early theoretical results which do not fit
later experimental data (see discussion in [48]).

\section{Discussion and Conclusions}

It is known that for positive atomic ions the Gaunt factor for transitions
with $\Delta n=0$ is larger then the Gaunt factor for transitions with $%
\Delta n>0\;$[15,10].{\sl \ }Here{\sl \ }we demonstrate similar regularity
for atoms.

With the Gaunt factor given by expressions (4) and (6), the Van Regemorter
formula{\sl \ }fits measured cross sections better than with the asymptotic
Gaunt factor (7) and semiempirical Gaunt factor [9]. This conclusion is
based on comparison with all available experimental cross sections for
allowed transitions $n\ell \rightarrow n^{\prime }\ell ^{\prime }\;$in atoms
with $LS$ coupling: a total of 23 cross sections for 11 atoms with various
electron configurations.

Inaccuracy of the Gaunt factor (4),(6) and inaccuracy of experimental data
are about the same. On one hand, good fit may be treated as a proof of weak
dependence of the Gaunt factor on electron configuration (for this class of
transitions). Then the Gaunt factor given by expressions (4) and (6)
provides acceptable accuracy of the cross section for any electron-atom
excitation (2),(3). This result is important for simulation of kinetic and
transport processes in low-temperature non-LTE plasmas (such as ionospheric,
gas-discharge, gas-laser, technological, or near-surface). On the other
hand, all studied transitions are from the states with $\ell =0\;$or $\ell =$
$1,$ therefore, for transitions from $d,f,g,.$.. states we have no proof of
independence of $G(x)$ on $\ell .$

Atoms of Neon, Argon, Krypton, and Xenon have $j\ell \;$coupling scheme.
Effective Gaunt factor $G_k^{\exp }\;$inferred from experimental cross
sections [55-58]{\sl \ } for these atoms, is a few times less than our Gaunt
factor (4),(6) which is quite accurate for atoms with $LS$ coupling. This
observation illustrates a dependence of the Gaunt factor on the coupling
scheme.

\section{Acknowledgments}

It is a pleasure to acknowledge fruitful discussions with Yuri Ralchenko. We
are grateful to National Institute for Fusion Science (Nagoya, Japan) and H.
Tawara for giving us an opportunity to use the NIFS atomic database. We are
also grateful to the database of Opacity Project. This work was supported by
the Israeli Academy of Science, Ministry of Science and Arts, and the
Ministry of Absorption.

\begin{center}
\ \ \ \ \ \ \ \ \ \ 

\ \ \ \ \ \ \ \ \ \ \ \ 

\_\_\_\_\_\_\_\_\_\_\_\_\_\_\_\_\_\_\_\_\_\_\_\_\_\_\_\_\_\_\_\_\_\_\_\_\_\_%
\_\_\_\_\_

\ \ \ \ 
\end{center}

[ 1] R. D. Cowan, {\it The Theory of Atomic Structure and Spectra}
(University of California Press, Berkeley, 1981)

[ 2] L. A. Vainshtein and V. P. Shevelko, {\it Structure and Features of
Ions in Hot Plasmas} (Nauka, Moscow, 1986) in Russian.

[ 3] V. P. Shevelko and L. A. Vainshtein, {\it Atomic Physics for Hot Plasmas%
} (IPP, Bristol, 1993).

[ 4] A. K. Pradhan, Atomic data and nuclear data tables, {\bf 52}, 227-317
(1992).

[ 5] D. Sampson and H. Zhang, Phys. Rev. A {\bf 45}, 1556 (1992).

[ 6]{\it \ Atomic and Molecular Processes in R-matrix Approach}, edited by
P. G. Burke and K. A. Berrington (IPP, Bristol, 1993).\ 

[ 7] A few atomic codes are available via computer networks, see, e.g., URL
http://plasma-gate.weizmann.ac.il/FSfAPP.html

[ 8] H. Van Regemorter, Astrophys. J. {\bf 136}, 906 (1962).

[ 9] I. I. Sobelman, L. A. Vainshtein, and E. A. Yukov, {\it Excitation of
atoms and Broadening og Spectral lines}, (Springer, 1981).

[10] D. H. Crandall. Ch. 4 in\ {\it Physics of Ion-ion and Electron-ion
Collisions}, edited by F. Brouillard and J. W. Mc. Gowan (Plenum Press, New
York, 1983).

[11] The Van Regemorter formula is derived from the Bethe formula [12,8,9]
which is obtained for optically allowed transitions.

[12] H. A. Bethe, Ann. Phys.,{\sl \ }Lpz. {\bf 5} 325 (1930).\ \ 

[13] I. I. Sobelman, {\it Atomic Spectra and Radiative Transitions, }%
(Springer, 1979).

[14] R. Mewe, Astron. \& Astrophys. {\bf 20}, 215 (1972).

[15] S. M. Younger and W. L. Wieze, J. Quant. Spectrosc. Radial. Transfer 
{\bf 22}, 161 (1979).

[16] D. H. Sampson and H. L. Zhang, Astroph. J, {\bf 335}, 516-524, (1988).

[17] H. Golten, to be submitted for publication.

[18] This criticism is caused by a discrepancy of a factor of 10 in
nearthreshold cross sections for some transitions $n,l,j$ $\rightarrow
n^{\prime }\ell ^{\prime }j^{\prime }\;$in multiple ions [5]. Probably, to
provide acceptable accuracy for multiple ions, the Gaunt factor has to
depend on more parameters of transition (for example, on $\ell $\ [17]) but
we do not consider multiple ions in this paper.

[19] I. Zapesochnyi, E. Postoi and I. Aleksakhin, Sov. Phys. JETP {\bf 41},
865 (1976).

[20] Database of National Institute for Fusion Science (Nagoya, Japan).

[21] D. Leep and A. Gallgher, Phys. Rev. A {\bf 10}, 1082 (1974).

[22] E. Enemark and A. Gallgher,Phys. Rev. A {\bf 6}, 192 (1972).

[23] D. Leep and A. Gallgher, Phys. Rev. A {\bf 13}, 148 (1976).

[24] W. Williams and S. Traymar, J. Phys. B {\bf 10}, 1955 (1977).

[25] S. Chen and A. Gallgher,Phys. Rev. A {\bf 14}, 593 (1976).

[26] R. Long, D. Cox and S. Smith, Journal of Research of the National
Bureau of Standarts.{\sl -}A. Physics and achemistry {\bf 72A}, 521 (1968).

[27] A. Mahan, A. Gallgher and S. Smith, Phys. Rev. A {\bf 13}, 156 (1976).

[28] J. de Jongh and J. Van Eck in {\it Electronic and Atomic Collision
(Abstracts of Papersof the IXth International Conference \ on\ the Physics
of Electronic and Atomic Collisions), }701 (1971).

[29] R. Hall,\ G.Joyez, J.Mazeau, J. Reinhardt and C.Schermann, Le Journal
de Physique {\bf 34}, 827 (1973).

[30] D. Shemansky, J. Ajello, D. Hall, and B. Franklin, The Astrophysical
Journal {\bf 296}, 774 (1985).

[31] G. Joyez, A. Huetz, M. Landau, J. Mazeau and F. Pishou, in {\it %
Electronic and Atomic Collision (Abstracts of Papersof the IXth\
International Conference on the Physics of Electronic and Atomic
Collisions), }827 (1975).

[32] M. Dillon and E. Lassettre, J. Chem. Phys. {\bf 62}, 2373 (1975).

[33] J. W. McConkey, F. G. Donaldson and M. A. Hander, Phys. Rev. Let.{\bf \
26}, 1413 (1971).

[34] F. Donaldson, M. Hender and J. McConkey, J. Phys. B {\bf 5}, 1192
(1972).

[35] W. Westerveld, H. Heideman and J. Van Eck, J. Phys. B {\bf 12 }115
(1979).

[36] H. Moustafa Mousa, F.de Heer, and J. Schutten, Physica, {\bf 40},517
(1969).

[37] D. Cartwring and G. Csanak, Phys. Rev. A 45, 1602 (1992).

[38] A. Chutjian and S. A. Sivastava, J. Phys.B {\bf 8}, 2360 (1975).

[39] J. Jobe and R. Jonh, Phys. Rev. {\bf 164}, 117 (1967).

[40] J. Showalter and R. Kay, Phys. Rev. A {\bf 11}, 1899 (1975).

[41] A. Van Raan, J. de Jongh, J. Van Eck and H. Heideman, Physica{\bf \ 53}%
, 45 (1971).

[42] I. Bogdanova and S. Urgenson, Opt. and Spectrosc. (USSR) {\bf 61}, 156
(1986).

[43] R.Jonh, F. Miller and C.Lin, Phys. Rev. {\bf 134 }A888 (1964).

[44] D. Rall F. A. Sharpton, M. B. Shulman, L. W. Anderson, J. E. Lawler and
C. C. Lin, Phys. Rev. Lett {\bf 62}, 2253 (1989).

[45] E. Zipf, J. Phys. B, {\bf 19} 2199 (1986).

[46] E.Stone and E.Zipf, J.Chem.Phys. {\bf 60}, 4237 (1974).

[47] S. Wang and J. McConkey, J. Phys. B {\bf 25}, 5461 (1992).

[48] J. Doering and L. Goembel,Journal of Geophysical Research, {\bf 96}%
,16,021(1991).

[49] A. A. Radzig and B. M. Smirnov {\it Parameters of Atoms and Atomic Ions 
}(Moscow, Energoatomizdat, 1986) in Russian.

[50] {\it The Opacity Project}. Volume 1. Institute of Physical Publishing,
Bristol and Philadelphia, 1995.

[51] Database of the Opacity Project (URL
http://cdsweb.u-strasbg.fr:80/topbase.html).

[52] W. L. Wiese, M. W. Smith, and B. M. Glennon {\it Atomic Transition
Probabilities.} Volume 1. National Standard Reference Data Series. National
Bureau of Standarts. Washington D.C., 1966.

[53] E. Stone and E. Zipf, J. Chem. Phys. {\bf 58}, 4278 (1973).

[54] D. Spence and P. Burrow, J. Phys.B {\bf 13}, 2809 (1980).

[55]{\sl \ }F. A. Sharton, R. M. St. John, C. C. Lin and F. E. Fajen, Phys.
Rev. A {\bf 2}, 1305 (1970).

[56] J. K. Ballou, C. C. Lin and F. E. Fajen, Phys. Rev. A {\bf 8}, 1797
(1973).

[57] R. S. Schappe, M. B. Shulman, L. W. Anderson and C. C. Lin, Phys. Rev.
A {\bf 50}, 444 (1994).

[58] I. P.{\sl \ }Zapesochnyi and P. V. Felstan,{\sl \ }Opt. and Spectrosk.
(USSR) {\bf 20,} 291 (1966)

\ \ \ 

\section{ Figure Captions}

{\bf Figure 1a. }The Gaunt factor for allowed transitions with $\Delta n=0$\
in atoms with $LS$ coupling. The values $G_k^{\exp }\;$are inferred from
experimental cross sections (listed in Table 1) using the Van Regemorter
formula.

{\bf Figure 1b. }Fragment of Figure 1a.

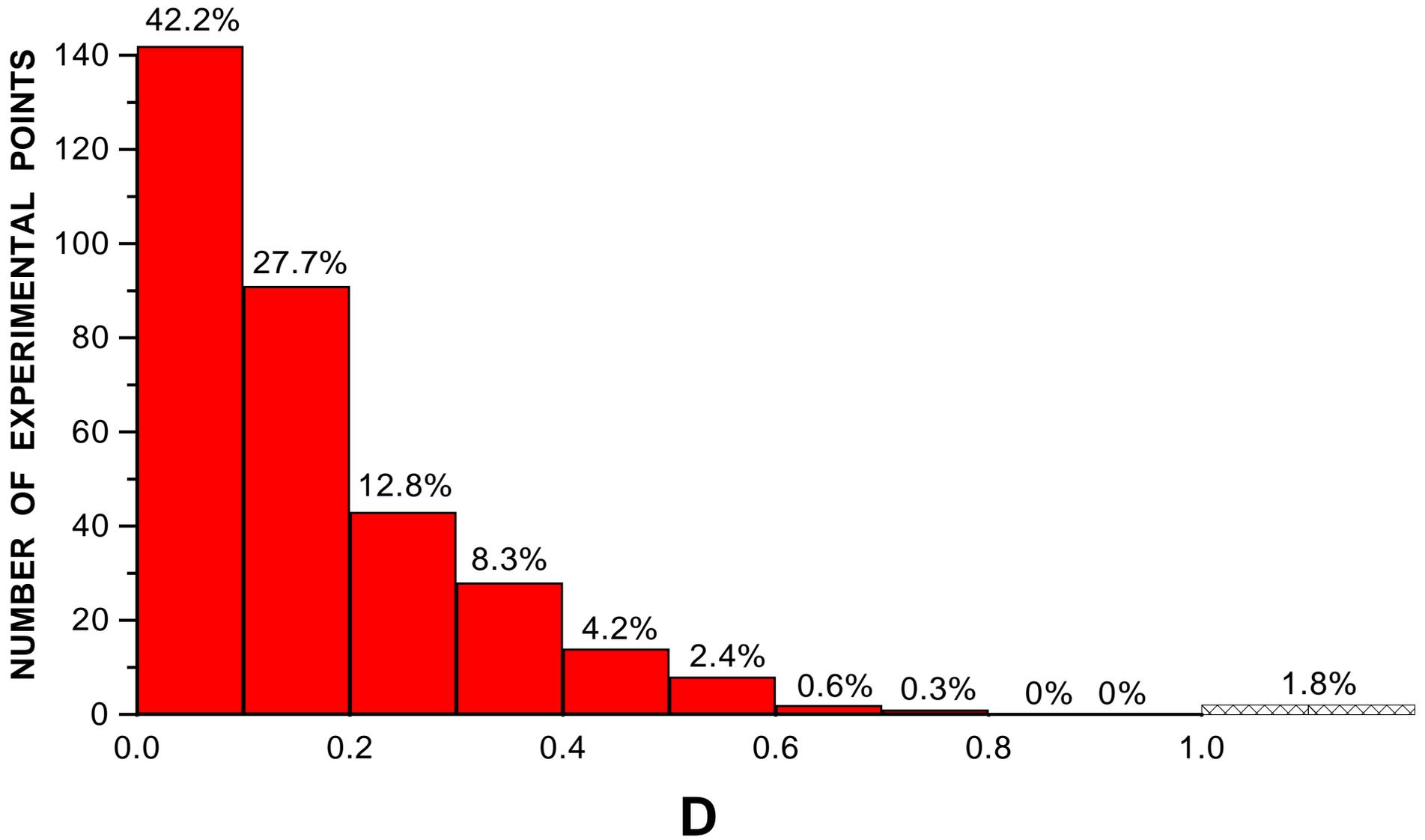
{\bf Figure 2. }Distribution of experimental values $G_k^{\exp }\;$presented
in Figure 1a over their deviation from the Gaunt factor (4): numbers of
experimental values of $G_k^{\exp }\;$per 10\% intervals of deviation (5).

{\bf Figure 3a. }The Gaunt factor for allowed transitions with $\Delta n>0$\
in atoms with $LS$ coupling. The values $G_k^{\exp }\;$are nferred from
experimental cross sections (listed in Table 2) using the Van Regemorter
formula.

{\bf Figure 3b. }Fragment of Figure 3a.

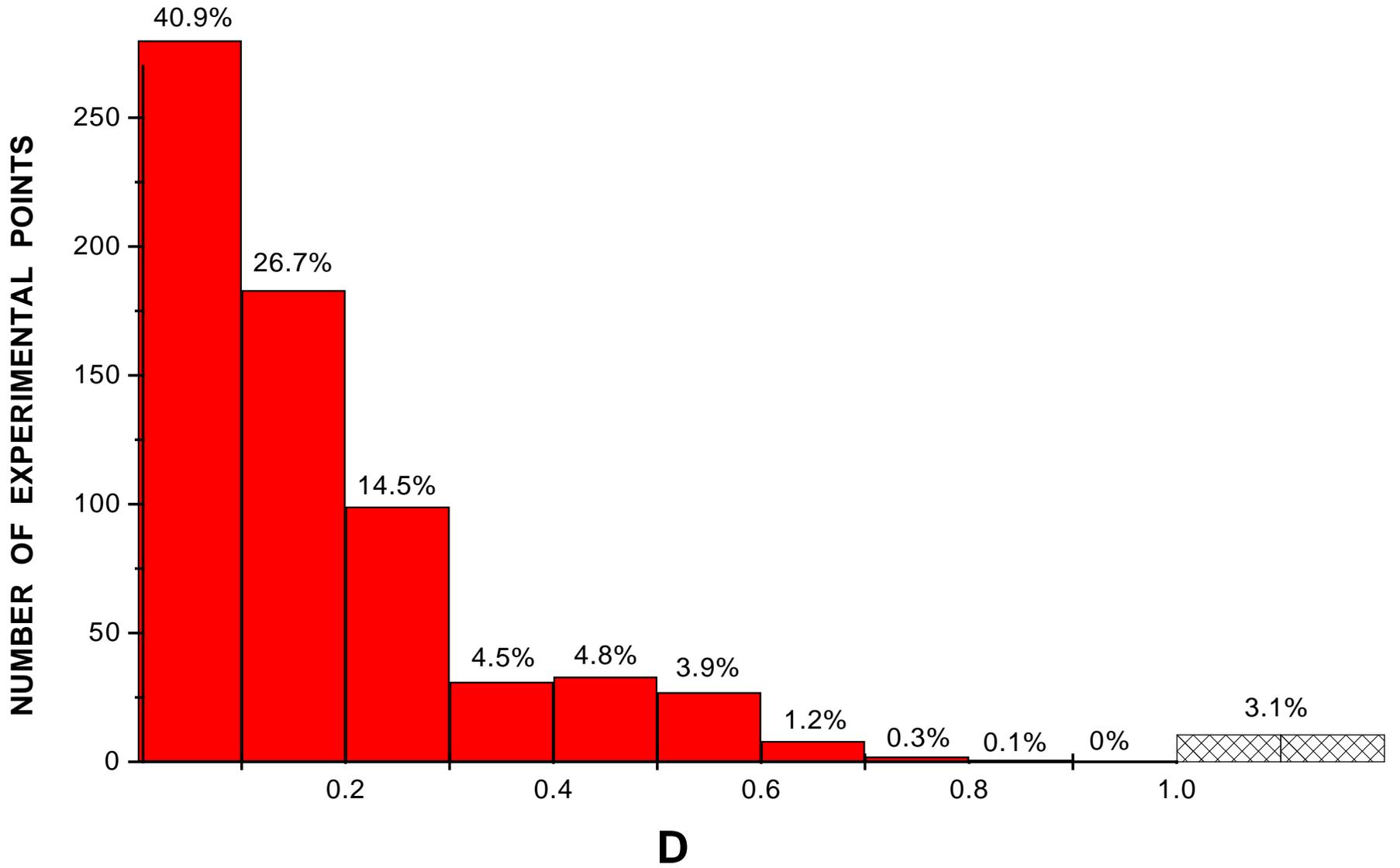
{\bf Figure 4. }Distribution of experimental values $G_k^{\exp }\;$presented
in Figure 3a over their deviation from the Gaunt factor (6): numbers of
experimental values of $G_k^{\exp }\;$per 10\% intervals of deviation $D_k=%
\frac{\mid G_k^{\exp }-\;G_{>}(x_k)\mid }{G_{>}(x_k)}$.

{\bf Figure 5a. }The Gaunt factor for transition ${\sl 3s-3p}$\ in $Na.$

{\bf Figure 5b. }Fragment of Figure 5a.

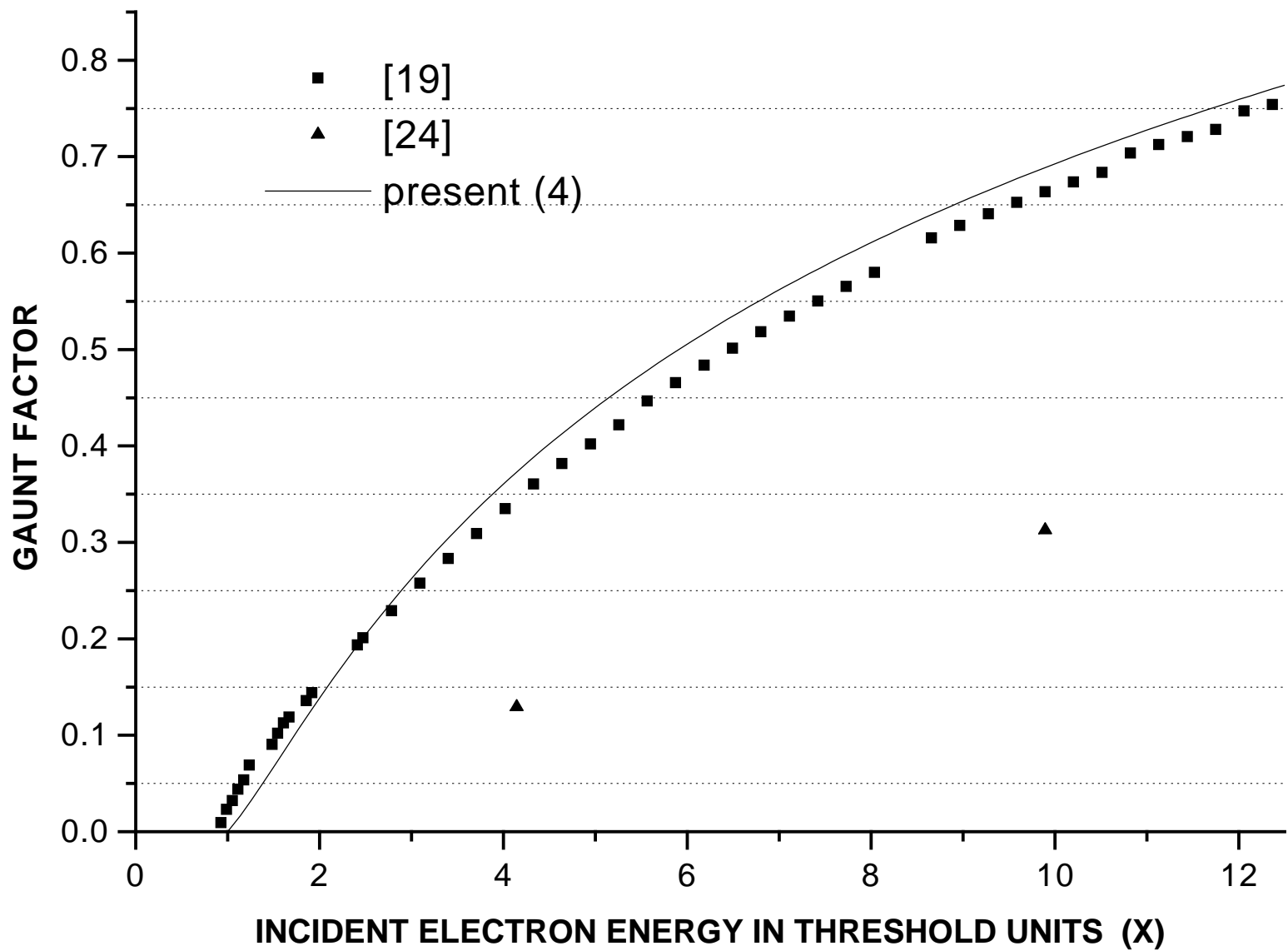
{\bf Figure 6. }The Gaunt factor for transition ${\sl 4s-4p}$\ in $K.$

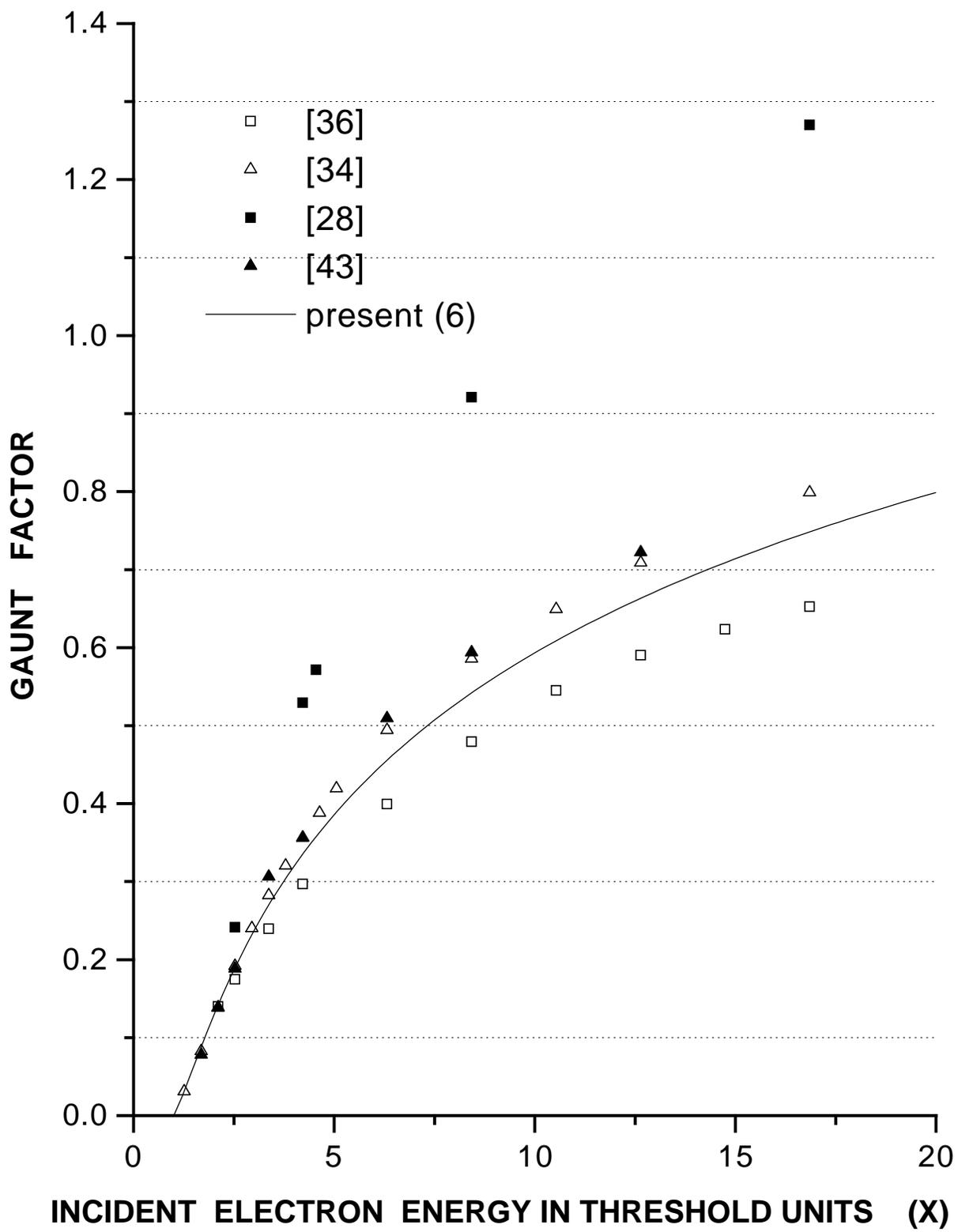
{\bf Figure 7. }The Gaunt factor for transition ${\sl 1s}^{2\;1}{\sl %
S\rightarrow 1s4p\;}^1{\sl P\;}$\ in $He.$

\section{Table Captions}

{\bf Table 1}. Experimentally studied allowed outer-shell transitions with $%
\Delta n=0$\ in atoms with $LS$ coupling.

{\bf Table 2}. Experimentally studied allowed outer-shell transitions with $%
\Delta n>0$\ in atoms with $LS$ coupling.

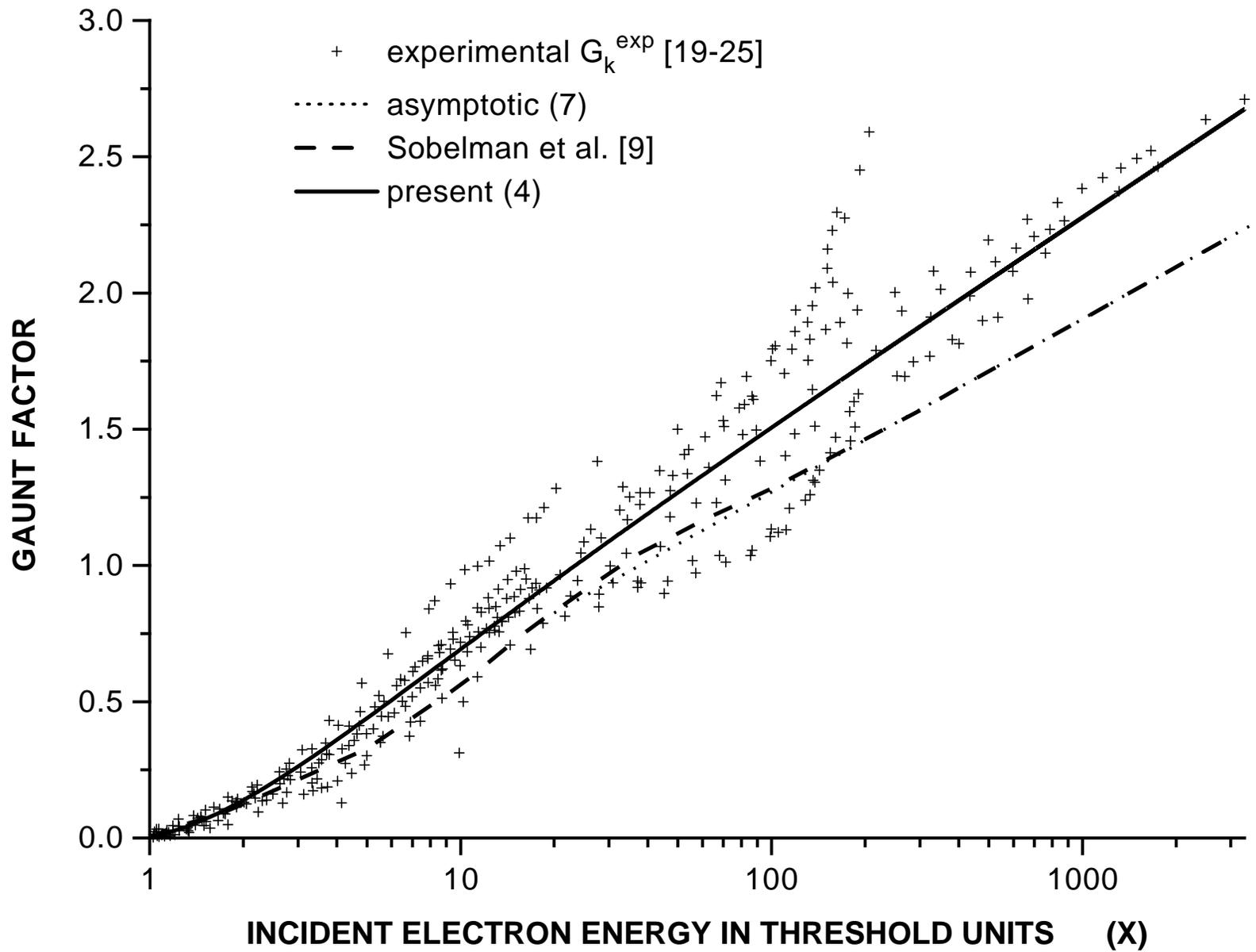
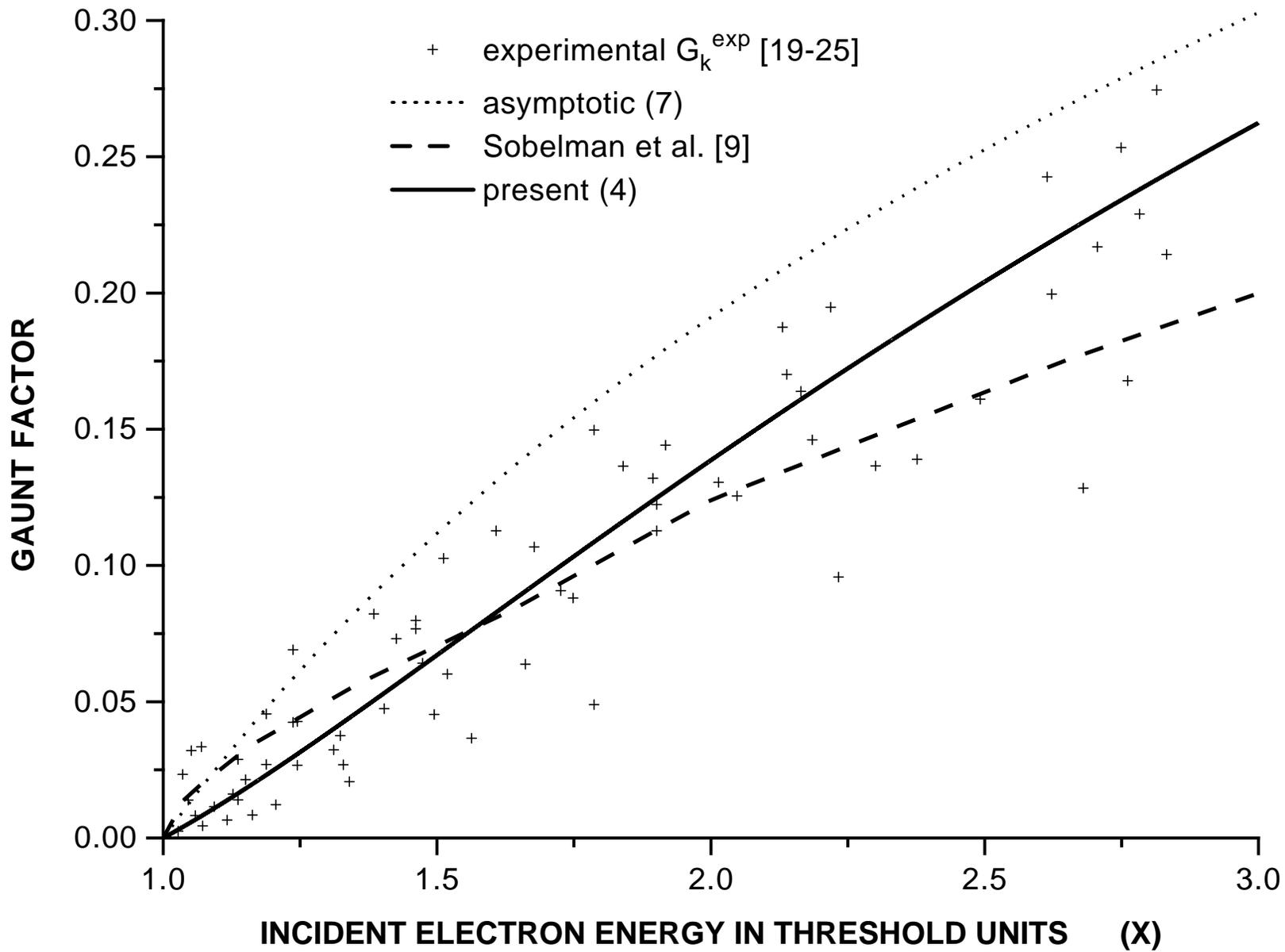
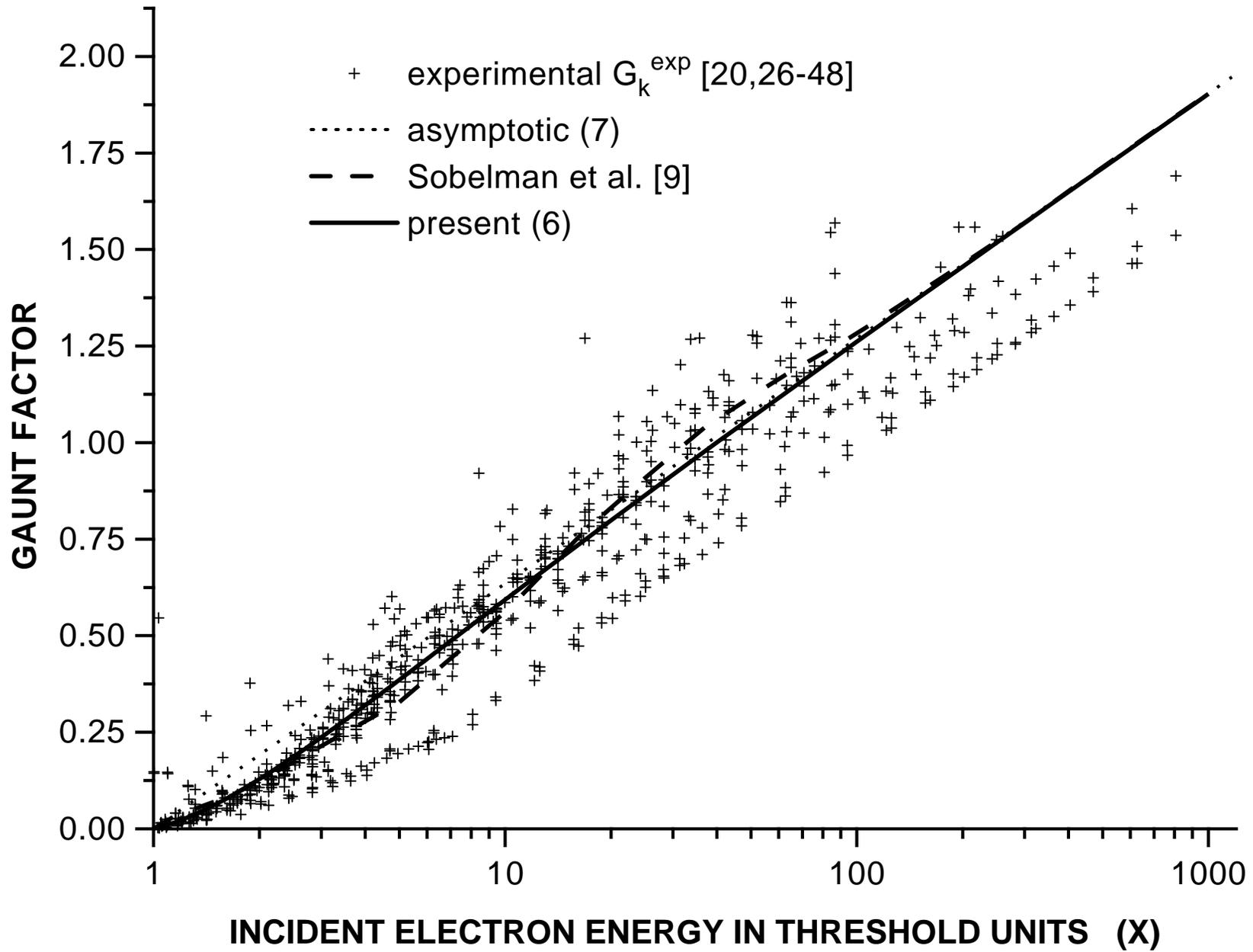
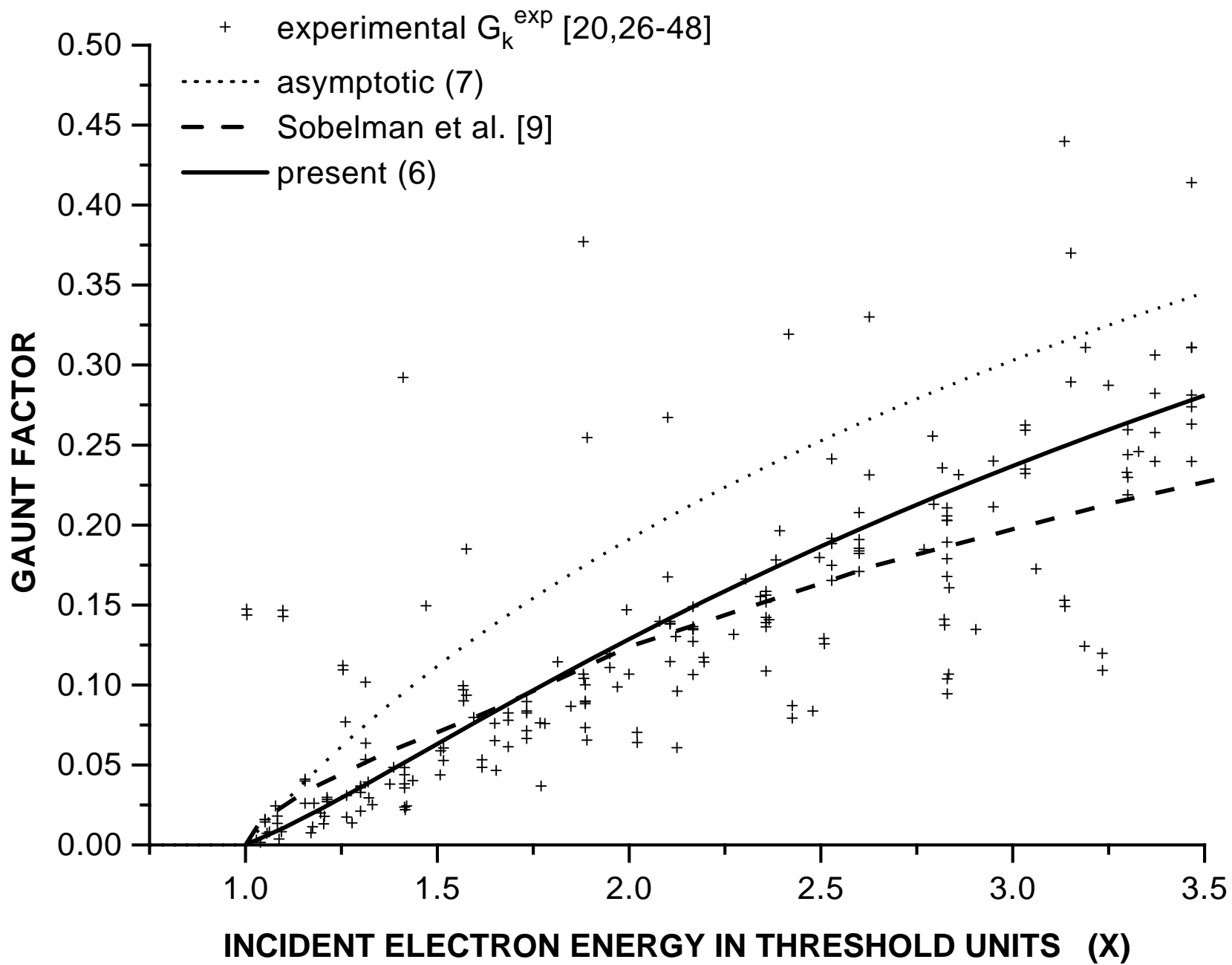
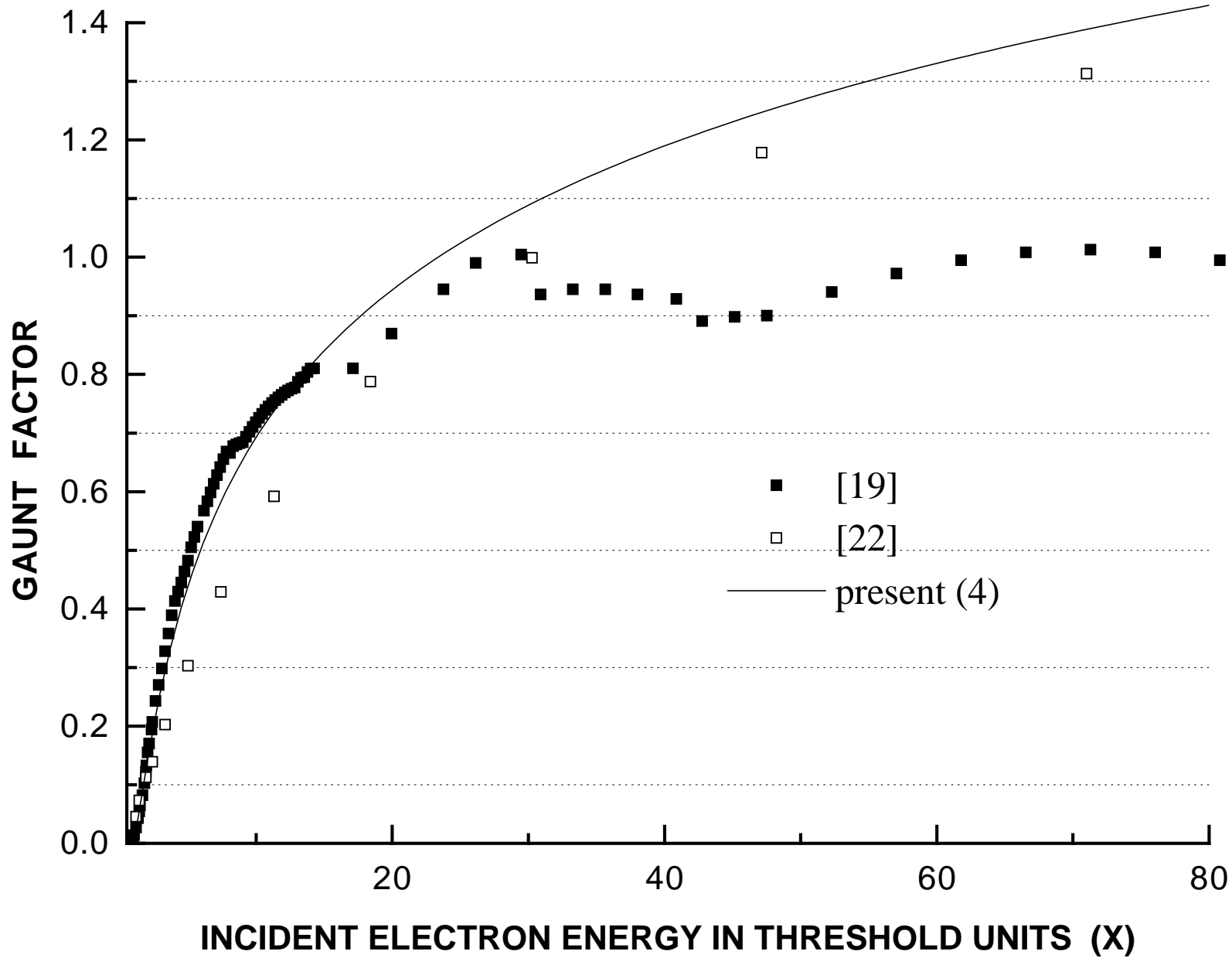
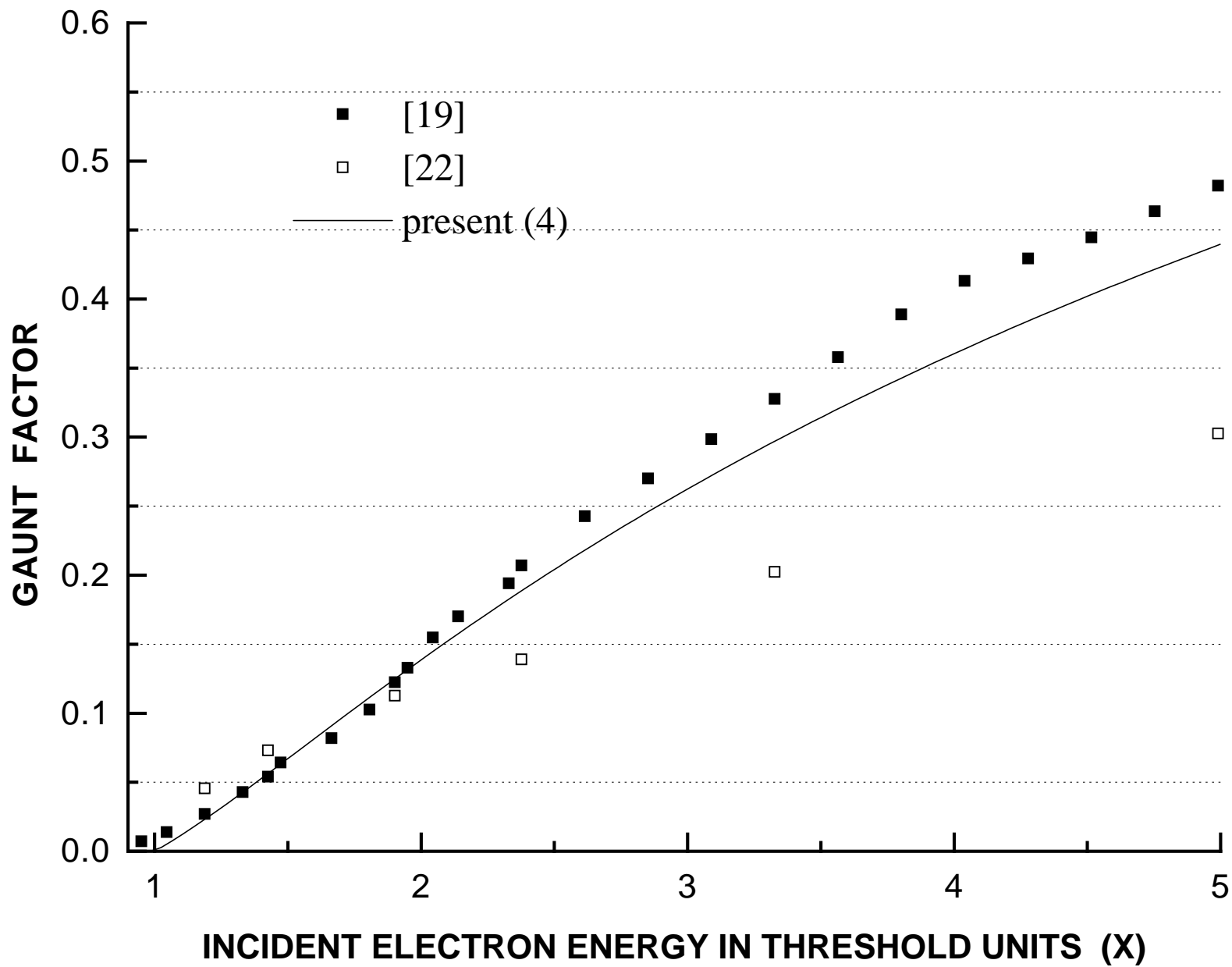
\end{document}